\documentclass[prc,twocolumn,showpacs,amsmath,amssymb]{revtex4}
\usepackage{times}
\usepackage{graphicx}
\usepackage{dcolumn}
\usepackage{bm}
\usepackage{amstext}

\begin{document}

\title{ Softness of Sn isotopes in relativistic semi-classical approximation}
\author{S. K. Biswal, S. K. Singh, M. Bhuyan and S. K. Patra}
\affiliation{
Institute of Physics, Sachivalaya Marg, Bhubaneswar-751 005, India. 
}
\date{\today}

\begin{abstract}
Within the frame-work of relativistic Thomas-Fermi and relativistic
extended Thomas-Fermi approximations, we calculate the giant monopole resonance 
(GMR) excitation energies
for Sn and related nuclei. A large number of non-linear relativistic
force parameters are used in this calculations. We find that a parameter set
is capable to reproduce the experimental monopole energy of Sn isotopes, 
when its nuclear matter compressibility lies within $210-230$ MeV, however
fails to reproduce the GMR energy of other related nuclei.
That means, simultaneously a parameter set can not
reproduce the GMR values of Sn and other nuclei.

\end{abstract}
\pacs{24.30.Cz, 21.10.Re, 24.10.Jv, 21.65.-f, 21.60.Ev, 21.65.Mn }
\maketitle

\section{Introduction}
Incompressibility of nuclear matter, also knows as compressional modulus has a 
special interest in pure nuclear and astro-nuclear physics, because of its 
fundamental role in guiding the equation of state (EOS) for nuclear matter.
Compressional modulus $K_{\infty}$ can not be measured by any experimental technique 
directly, rather it depends indirectly on the experimental measurement of 
isoscalar giant monopole resonance (ISGMR) for its conformation \cite {blaizot}. 
This fact enriches the demand of correct measurement of excitation energy of 
ISGMR. The relativistic parameter with random phase approaches (RPA) constraint the 
range of the compressibility modulus $270\pm{10}$ MeV \cite {lala97,Vrete02} for 
nuclear matter. Similarly, the non-relativistic formalism with Hartree-Fock 
(HF) plus RPA allows the acceptable range of compressibility modulus 
$210-220$ MeV, which is less than relativistic one. 
It is believed that the part of this discrepancy in the acceptable range of 
compressional modulus comes from the diverse behavior of the density 
dependence of symmetry energy in relativistic and non-relativistic formalism 
\cite {horow02,horowr01}. Now, both the relativistic and non-relativistic formalisms
come with a general agreement on the value of nuclear compressibility 
i.e., $240\pm{10}$ MeV \cite{colo04,todd05,Agrawal05}. But the new experiment 
on $Sn$ isotopic series i.e., $^{112}Sn-^{124}Sn$ rises the question 
"why Tin is so fluffy ?"\cite {ugarg,tli07,ugarg07}. This question again finger 
towards the correct theoretical investigation of compressibility modulus. 

Thus, it is worthy to investigate the compressibility modulus 
in various theoretical formalisms. Most of the relativistic and non-relativistic
theoretical models reproduce the strength distribution very well for medium 
and heavy nuclei, like $^{90}Zr$ and $^{208}{Pb}$, respectively. But at the same 
time it overestimate the excitation energy of $Sn$ around 1 MeV. This low value
 of excitation energy demands lower value nuclear matter compressibility. 
This gives a new challenge to both the theoretical and experimental nuclear 
physicsts. Till date, lots of effort have been devoted to solve this problem like 
inclusion of pairing effect \cite {citav91,gang12,jun08,khan09}, mutually 
enhancement effect (MEM) etc. \cite {khana09}. But the pairing effect reduces 
the theoretical excitation energy only by 150 KeV in $Sn$ isotopic 
series, which may not be sufficient to overcome the puzzle. Similarly, 
new experimental data are not in favor of MEM effect\cite{dpatel13}. Measurement
on excitation energy of $^{204,206,208}Pb$ shows that the MEM effect should 
rule out in manifestation of stiffness of the $Sn$ isotopic series. 

Here, in the present work, we use the relativistic Thomas-Fermi (RTF) and  
relativistic extended Thomas-Fermi (RETF) 
\cite{acentel93,acentel92,speich93,centel98,centell93} 
with scaling and constraint approaches in the frame-work of non-linear 
$\sigma-\omega$ model \cite {bo77}. The RETF is the $\hbar^2$ correction 
to the RTF, where variation of density taken care properly mostly in the
surface of the nucleus \cite{cetelles90}. The RETF formalism is 
more towards the quantal Hartee approximation. It is also verified that 
the semiclassical approximation like Thomas-Fermi method is very useful in 
calculation of collective property of nucleus, like giant monopole resonance
(GMR). In particular, for heavier mass nuclei, it gives almost similar results
with the complicated quantal calculation. This is because of quantal 
correction are averaged out in heavier mass nuclei and results are inclined toward 
semiclassical one. So it is very much instructive to calculate excitation energy 
of various nucleus by this method and compared with experimental results. 
Since, last one decade, the softness of Sn isotopes remain a headache for both 
theorists and experimentalists, it is worthy to discuss the softness of Tin 
isotopes in semiclassical approximations like RETF and RTF. 

Here, we calculate the excitation energy of 
Sn isotopes from $^{112}$Sn to $^{124}$Sn using the semi-classical
RTF and RETF model. The different 
momentum ratio like $(m_3/m_1)^{1/2}$ and $(m_1/m_{-1})^{1/2}$ 
are compared with the scaling and constraint calculations. The 
theoretical results are computed in various parameter sets such as
NL-SH, NL1, NL2, Nl3 and FSUG. We analyzed 
the predictive power of these parameter sets and discussed
various aspects of the compressibility modulus.

This paper is organized as follow: In Section II we have 
summarized the theoretical formalisms, which are useful
for the present analysis. In section III, we have given 
the discussions of our results. Here, we have elaborated
the giant monopole resonance obtained by various parameter sets
and their connectivity with compressibility modulus. The
last section is devoted to a summary and concluding
remarks.

\section{Theoretical framework}
The principle of scale invariance is used to obtain 
the virial theorem for the relativistic mean field \cite{serot86} theory 
by working in the relativistic Thomas--Fermi (RTF) and relativistic 
extended Thomas-Fermi (RETF) approximations 
\cite{patra01,acentel93,acentel92,spei98,mario98,mario93a}. 
Although, the scaling and constrained calculations 
are not new, the present technique is developed first time by Patra et 
al \cite{patra01,mario10}, which is different from other scaling formalisms.

The detail formalisms of 
the scaling method are given in Refs. \cite{patra01,patra02}. For 
completeness, we have outlined briefly some of the essential expressions, 
which are needed for the present purpose. We have worked  with the 
non-linear Lagrangian of Boguta and Bodmer \cite{bo77} to include 
the many-body correlation arises from the non-linear 
terms of the $\sigma-$meson self-interaction \cite{schiff51,fujita57,steven01}
for nuclear many-body system. The nuclear matter compressibility modulus 
$K_{\infty}$ also reduces dramatically by the introduction of these terms, 
which motivates to work with this non-liner Lagrangian. The relativistic mean 
field Hamiltonian for a nucleon-meson interacting system is written by 
\cite{serot86,patra01}:
\begin{eqnarray}
{\cal H}&= &\sum_i \varphi_i^{\dagger}
\bigg[ - i \vec{\alpha} \cdot \vec{\nabla} +
\beta m^* + g_{v} V + \frac{1}{2} g_{\rho} R \tau_3 \nonumber\\
&+&\frac{1}{2} e {\cal A} (1+\tau_3) \bigg] \varphi_i 
+ \frac{1}{2} \left[ (\vec{\nabla}\phi)^2 + m_{s}^2 \phi^2 \right]
+\frac{1}{3} b \phi^3\nonumber\\
&+& \frac{1}{4} c \phi^4
-\frac{1}{2} \left[ (\vec{\nabla} V)^2 + m_{v}^2 V^2 \right]\nonumber\\
&-& \frac{1}{2} \left[ (\vec{\nabla} R)^2 + m_\rho^2 R^2 \right]
- \frac{1}{2} \left(\vec{\nabla}  {\cal A}\right)^2 -
\frac {\zeta_0}{24} g_{v}^4{V}^4 \\ 
&-& {\Lambda_V} {g_v^2}{g_\rho}^2{R^2}{V^2} 
\end{eqnarray}
Here $m$, $m_s$, $m_v$ and $m_{\rho}$ are the masses for the
nucleon (with $m^*=m-g_s\phi$ being the effective mass of the nucleon),
${\sigma}-$, $\omega-$ and ${\rho}-$mesons, respectively
and ${\varphi}$ is the Dirac spinor.
The field for the ${\sigma}$-meson is denoted by ${\phi}$, for ${\omega}$-meson
by $V$, for ${\rho}$-meson by $R$ ($\tau_3$ as the $3^{rd}$ component of the
isospin)  and for photon by $A$.
$g_s$, $g_v$, $g_{\rho}$ and $e^2/4{\pi}$=1/137 are the coupling
constants for the ${\sigma}$, ${\omega}$, ${\rho}$-mesons and photon respectively.
$b$ and $c$ are the non-linear coupling constants for ${\sigma}$ mesons.
By using the classical variational principle we obtain the field equations for
the nucleon and mesons. In semi-classical approximation, we can write the above
Hamiltonian in term of density as:
\begin{eqnarray}
{\cal H}&=&{\cal E}+g_v V {\rho}+g_{\rho}R{\rho}_3+e{\cal A}{\rho}_p+{\cal H}_f,
\end{eqnarray}
where
\begin{eqnarray}
{\cal E}&= &\sum_i \varphi_i^{\dagger}
\bigg[ - i \vec{\alpha} \cdot \vec{\nabla} +
\beta m^*\bigg]\varphi_i,
\end{eqnarray}
\begin{eqnarray}
{\rho}_s&=&\sum_i \varphi_i^{\dagger}{\varphi},
\end{eqnarray}
\begin{eqnarray}
{\rho}&=&\sum_i {\bar \varphi_i}{\varphi},
\end{eqnarray}
\begin{eqnarray}
{\rho_3}&=& \frac {1}{2}\sum_i{\varphi}_i^{\dagger}{\tau_3}{\varphi}_i,
\end{eqnarray}
and ${\cal H}_f$ is the free part of the Hamiltonian.
The total density $\rho$ is the sum of proton $\rho_p$ and neutron
$\rho_n$ densities.
The semi-classical ground-state meson fields are obtained
by solving the Euler--Lagrange equations $\delta {\cal H}/\delta \rho_q = \mu_q$
($q=n, p$).
\begin{equation}
(\Delta- m_s^2)\phi = -g_{s} \rho_{s} +b\phi^2 +c\phi^3 ,
\label{eqFN4}  \\[3mm]
\end{equation}
\begin{eqnarray}
(\Delta - m_{v}^2) V &=&  -g_{v} \rho+ 2{\Lambda_V}{R^2}V \\
&+&\frac{\zeta_0}{6}{g_v^4}V^3 
\label{eq18} 
\end{eqnarray}
\begin{equation}
(\Delta - m_\rho^2)  R  =   - g_\rho \rho_3+2 {\Lambda_V}R{V^2},
\label{eq19} 
\end{equation}
\begin{equation}
\Delta {\cal A}    =      -e \rho_{p}.
\label{eqFN7}
\end{equation}
The above field equations are solved self-consistently in an iterative method.
\begin{multline}
{\cal H} = {\cal E} + \frac{1}{2}g_{s}\phi \rho^{eff}_{s}
+ \frac{1}{3}b\phi^3+\frac{1}{4} c \phi^4
+\frac{1}{2} g_{v}  V \rho +\frac{1}{2} g_\rho R \rho_3 
+\frac{1}{2} e {\cal A} \rho_{p}\\
-2{\Lambda_V}{R^2}{V^2}-\frac{\zeta_0}{12}{g_v}^4{V^4},
\label{eqFN8c}
\end{multline}
with
\begin {eqnarray}
{\rho}_s^{eff}&= & g_s{\rho}_s-b{\phi}^2-c{\phi}^3.
\end{eqnarray}
In order to study the monopole vibration of the nucleus we have scaled the baryon density
\cite{patra01}.
The normalized form of the baryon density is given by
\begin {eqnarray}
{\rho}_{\lambda}\left(\bf r \right)&=&{\lambda}^3{\rho}\left(\lambda r \right),
\end{eqnarray}
${\lambda}$ is the collective co-ordinate associated
 with the monopole vibration. As Fermi momentum and
 density are inter-related, the scaled Fermi momentum is given by
\begin {eqnarray}
K_Fq{\lambda}&=&\left[3{\pi}^2{\rho}_q
\lambda\left(\bf r \right)\right]^{\frac{1}{3}}.
\end {eqnarray}
Similarly $\phi$, $V$, $R$ and Coulomb fields are scaled due to self-consistence
eqs. (7-10).
But the $\phi$ field can not be scaled simply like the density and momentum,
 because the source term of $\phi $  field contains the $\phi$ field itself. In
semi-classical formalism, the energy and density are scaled like
\begin{eqnarray}
{\cal E_{\lambda}}(\bf r)&=&{\lambda}^4 {\tilde{\cal E}}(\lambda \bf r)\nonumber\\
&=& \lambda^4[\tilde{\cal E}_{0}(\lambda \bf r)+
\tilde{\cal E}_2(\lambda \bf r)],
\end {eqnarray}
\begin{eqnarray}
\rho_{{s\lambda}}(\bf r)={\lambda^3}{\tilde{\rho}}_{s}{(\lambda{\bf r})}.
\end{eqnarray}
The symbol $\sim$ shows an implicit dependence of $ \tilde m^* $.
With all these scaled variables, we can write the Hamiltonian as:
\begin{eqnarray}
{\cal{H}}_{\lambda} & = & {\lambda^3}{\lambda}{\tilde{\cal{E}}}+\frac{1}{2}{g_s}
\phi_{\lambda}{{\tilde{\rho}}_s^{eff}}\nonumber
+\frac{1}{3}\frac{b}{\lambda^3}\phi_\lambda^3
+\frac{1}{4}\frac{c}{\lambda^3}{\phi_\lambda}^4\\
&+&\frac{1}{2}{g_v}V_\lambda{\rho}
+\frac{1}{2}g_{\rho}R_\lambda {\rho_3}+\frac{1}{2}e{A}_\lambda{\rho}_p\\
&-& 2\Lambda_V {R_\lambda}^2{V_\lambda}^2 -\frac{\zeta_0}{12}{g_v^4}{
V_\lambda}^4
\end{eqnarray}
Here we are interested to calculate the monopole excitation energy which is defined as
${E}^{s}={\sqrt{\frac{C_m}{B_m}}}$ with ${C_m}$ is the restoring force and
$B_m$ is the mass parameter. In our calculations, $C_m$ is
obtained from the double derivative of the scaled energy with respect to the scaled
co-ordinate $\lambda$ at ${\lambda}=1$ and is defined as \cite{patra01}:
\begin {eqnarray}
{C}_m&=&\int{dr}\bigg[-m\frac{\partial{\tilde{\rho_s}}}{\partial{\lambda}}
+3\bigg({m_s}^2{\phi}^2+\frac{1}{3}b{\phi}^3
\nonumber\\
&-& {m_v}^2{V^2}
-{m_{\rho}}^2R^2\bigg)-(2{m_s}^2{\phi}
+b{\phi}^2)\frac{\partial{\phi_\lambda}}{\partial{\lambda}}\nonumber\\
&+&2{m_v}^2V\frac{\partial{V_\lambda}}{\partial{\lambda}}
+2{m_{\rho}}^2R
\frac{\partial R_\lambda}{\partial \lambda}\bigg]_{\lambda=1},
\end{eqnarray}
and the mass parameter $B_{m}$ of the monopole vibration can be expressed as the
double derivative of the scaled energy with the collective velocity $\dot{\lambda}$
as
\begin{eqnarray}
B_{m}=\int{dr}{U(\bf r)}^2{\cal {H}},
\end {eqnarray}
where $U(\bf r)$ is the displacement field, which can be determined from the relation
between collective velocity $\dot{\lambda}$ and velocity of the moving frame,
\begin {eqnarray}
U(\bf r)=\frac{1}{\rho(\bf r){\bf r}^2}\int{dr'}{\rho}_T(r'){r'}^2,
\end {eqnarray}
with ${\rho}_T $ is the transition density defined as
\begin {eqnarray}
{{\rho}_T(\bf r)}=\frac{\partial{\rho_\lambda(\bf r)}}{\partial{\lambda}}\bigg|_{\lambda = 1}
=3 {\rho}(\bf r)+r \frac{\partial{\rho(\bf r)}}{\partial r},
\end {eqnarray}
taking $U(\bf r)=r $. Then the mass parameter can be written as
$B_m=\int{dr}{r}^2{\cal H}$.
In non-relativistic limit, ${B_m}^{nr}=\int{dr}{r^2}m{\rho}$ and the
scaled energy ${E_m}^{s}$ is $\sqrt{\frac{m_3}{m_1}}$.
The expressions for ${m_3}$ and ${m_1}$
can be found in \cite {bohigas79}.
Along with the scaling calculation, the monopole vibration can
also be studied with constrained approach \cite {bohigas79,maru89,boer91,stoi94,stoi94a}.
In this method, one has to solve the
constrained functional equation:
\begin{eqnarray}
\int{dr}\left[{\cal H}-{\eta}{r}^2 {\rho}\right]=E(\eta)-\eta\int{dr}{r}^2\rho.
\end{eqnarray}
Here the constrained is ${\langle {R^2}\rangle}_0 ={\langle{r^2}\rangle}_m$. The
constrained energy $E(\eta)$ can be expanded in a harmonic approximation as
\begin{eqnarray}
E(\eta)& = &E(0)+\frac{\partial E(\eta)}{\partial \eta}\big|_{\eta =0}
+\frac{\partial^2{E(\eta)}}{\partial{\eta}^2}|_{\eta =0}.
\end {eqnarray}
The second order derivative in the expansion is related with the constrained
compressibility modulus for finite nucleus $K_A^c$ as
\begin{eqnarray}
{K_A}^{c} = \frac{1}{A} {R_0}^{2} \frac{\partial^2{E \eta}}
{\partial{R_\eta}},
\end{eqnarray}
and the constrained energy ${{E_m}^{c}}$ as
\begin {eqnarray}
{{E_m}^{c}}={\sqrt{\frac{A {K_A^c}}{B_m^c}}}.
\end {eqnarray}
In the non-relativistic approach, the constrained energy is related by the
sum rule weighted ${{E_m}^{c}}={\sqrt{\frac{m_1}{m_{-1}}}}$. Now the scaling and constrained
excitation energies of the monopole vibration in terms of the non-relativistic
sum rules will help us to calculate $\sigma$, i.e. the resonance width 
\cite{bohigas79,centelle05},
\begin{eqnarray}
{\sigma} &=& \sqrt{\left({E_m}^s\right)^2-\left({E_m}^c\right)^2} 
= \sqrt{({\frac{m_3}{m_1}})^2-{(\frac{m_1}{m_{-1}}})^2}. 
\end{eqnarray}






\section{Results and Discussions}

It is interesting to apply the model to calculate the excitation energy of Sn 
isotopic series and compared  with experimental results. 
Thus, we calculate the GMR energy using both the scaling and constraint 
methods in the frame-work of relativistic
extended Thomas-Fermi approximation using various parameter sets
for Z= 48 and 50 and compared with 
the excitation energy with momentum ratio $m_3/m_1$ and $m_1/m_{-1}$ obtained  
from multipule decomposition analysis (MDA). 
The basic reason 
to take a number of parameter sets is that the infinite nuclear matter 
compressibility of 
these forces cover a wide range of values. For example, NL-SH has 
compressibility 399 MeV, while that of NL1 is 210 MeV. From MDA analysis we 
get different momentum ratio, such as $m_3/m_1$, $m_0/m_1$ and $m_1/m_{-1}$.
These ratios are connected to scaling, centroid and constraint 
energies, respectively. 
That is why we compared our theoretical scaling result 
with $({m_3/m_1})^{1/2}$ and $({m_1/m_{-1}})^{1/2}$ with the constrained 
calculations. 
\begin{figure}
\vspace{0.6cm}
\hspace{-0.3cm}
\includegraphics[scale=0.32]{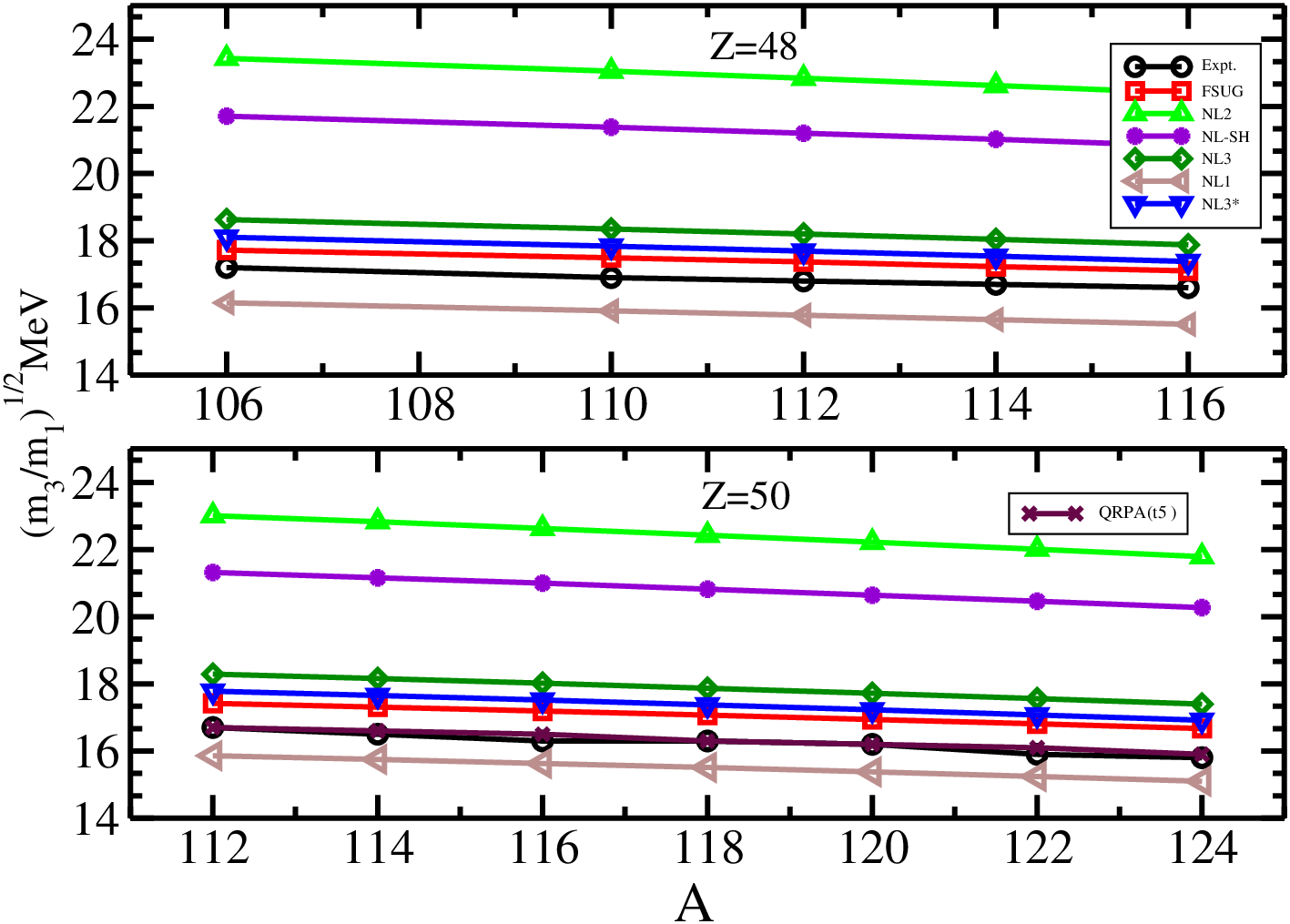}
\caption{\label{fig1} Giant monopole excitation energy obtained 
by scaling method with various parameter sets are compared with 
experimental \cite{ugarg,tli07,ugarg07} $({m_3/m_1})^{1/2}$ for 
Cd and Sn isotopes.}
\end{figure}

\begin{figure}
\vspace{0.6cm}
\hspace{-0.3cm}
\includegraphics[scale=0.32]{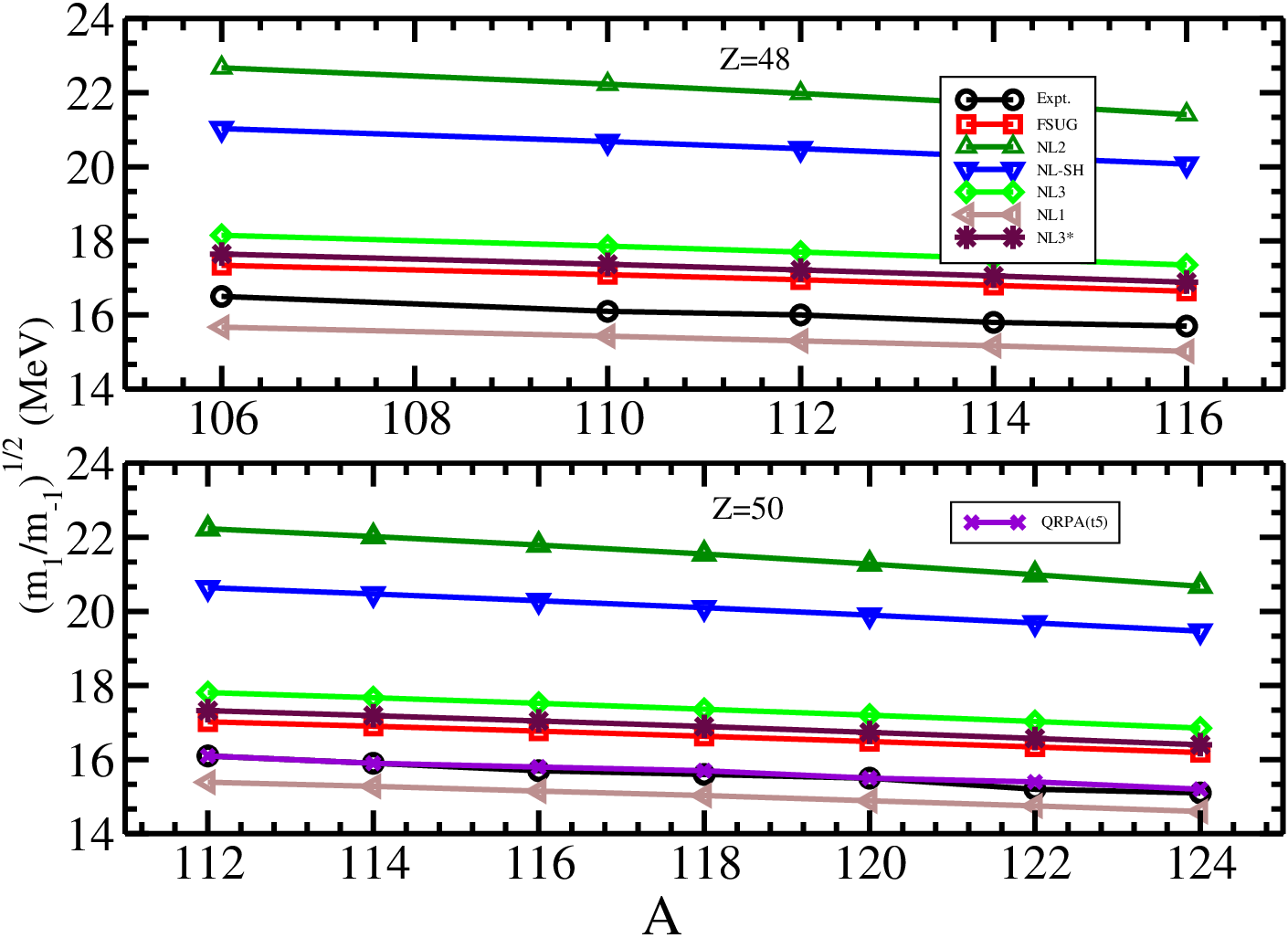}
\caption{\label{fig2} Same as Fig. 1, but for $({m_1/m_{-1}})^{1/2}$.}
\end{figure}

In Figures 1 and 2 we have shown the $(m_3/m_1)^{1/2}$ and 
$(m_1/m_{-1})^{1/2}$ ratio for isotopic chains of Cd and Sn. The results are 
also compared with experimental data obtained from (RCNP) \cite 
{ugarg,tli07,ugarg07}. From 
the figures, it is cleared that the experimental value lies between the 
results obtained from FUSG (FSUGold) and NL1  force parameters. It is to be
noted that, throughout the calculations, we have used only the  
non-linner parameter sets for their excellent prediction of nuclear
observables with the experimental data. If one compares the experimental 
and theoretical results for $^{208}$Pb, the FSUG set gives better
results amongst all. For example, the experimental and theoretical 
data are 14.17$\pm$0.1 and 14.04 MeV, respectively. These values are 
well matched with each other.
From this, one could conclude that the infinite 
nuclear matter compressibility lies nearer to that of FSUG (230.28 MeV) 
parameter. But experimental result on ISGM in RCNP shows that the predictive 
power of FUSG is not good enough for the excitation energy of Sn 
isotopes. This observation is not only confined to RCNP formalism, 
but also persists in the more sophisticated RPA approach. 
In Table I,  we have given the 
results for QRPA(T6), RETF(FSUG) and RETF(NL1). The experimental data are also
given to compare all these theoretical results.

\bigskip
\begin{table*}
\caption{\label{table1}{Momentum ratio for Sn isotopes using RETF
approximation with FSUGold and NL1 sets are compared with QRPA(T6)
predictions \cite{tsel09}.}}
\bigskip
\begin{tabular}{|c|c|c|c|c|c|c|c|c|}
\hline
Nucleus &\multicolumn{4}{c|}{$({m_3/m_1})^{1/2}$(MeV)}&\multicolumn{4}{c|}{$({m_1/m_{-1}})^{1/2}$(MeV)}\\ \hline
 &QRPA(T6)&RETF(FSUG)&RETF(NL1)& Expt. & QRPA(T6)&RETF(FSU)&RETF(NL1)& Expt. \\ \hline
$^{112}$Sn&17.3& 17.42&15.86 &16.7& 17.0&17.2&15.39&16.1\\
$^{114}$Sn&17.2& 17.32&15.75 &16.5& 16.9&16.9&15.28&15.9 \\
$^{116}$Sn&17.1& 17.19&15.63&16.3&16.8&16.77&15.15&15.7 \\
$^{118}$Sn&17.0&17.07&15.51&16.3&16.6&16.63&15.03&15.6 \\
$^{120}$Sn&16.9&16.94&15.38 &16.2& 16.5&16.44&14.89&15.5 \\
$^{122}$Sn& 16.8&16.81 &15.24 &15.9& 16.4& 16.34&14.75&15.2\\
$^{124}$Sn&16.7&16.67&15.1&15.8&16.2&16.19&14.6&15.1\\
\hline
\end{tabular}
\end{table*}
The infinite nuclear matter compressibility $K_{\infty}$ with T6 parameter set is 
236 MeV and that of FSUG is 230.28 MeV. The difference in $K_{\infty}$ 
between these two sets is only 6 MeV. The similarity in compressibility 
(small difference in $K_{\infty}$)  may be a reason for their prediction
in equal value of GMR. The table shows that, there is only 0.1 MeV difference 
in QRPA(T5) and RETF(FSU) results in the GMR values for 
$^{112}$Sn$-$$^{116}$Sn  isotopes, but the results are exactly matched 
for the $^{118}$Sn$-$$^{124}$Sn. This implies 
that for relatively higher mass nuclei, both the QRPA(T6) and RETF(FSUG) 
results are almost similar. If some one consider the experimental value 
of Sn isotopic series, then QRPA(T5) gives better result. For example,
experimental value of $(m_3/m_1)^{1/2}$ for $^{112}$Sn is 16.7$\pm$ 0.2 MeV 
and that for QRPA(T5) is 16.6 MeV. These two values matches well with
each other. 
The infinite nuclear matter compressibility of T5 set is 202 MeV. It is
shown by  V. Tselyaev et al. \cite{tsel09} that the T5 parameter set 
with such compressibility, better explains the excitation energy of Sn 
isotopes, but fails to predict the excitation energy of $^{208}$Pb.
It over estimates the data for $^{208}$Pb. The experimental data of 
ISGMR energies for $^{90}$Zr and $^{114}$Sn lies in between the calculated 
values of T5 and T6 forces. 
In summary, we can say that the RPA analysis predicts the 
symmetric nuclear matter compressibility within $202-236$ MeV and our 
semi-classical calculation gives it in the  range $210-230$ MeV.
These two predictions almost agree with each other in the acceptable
limit.

\begin{table}
\caption{\label{table2}{Momentum ratio ${\sqrt{{m_1/m_{-1}}}}$ for Pb 
isotopes within  RETF is compared  with pairing+ MEM results and 
experimental data \cite{dpatel13}.}}
\bigskip
\begin{tabular}{|c|c|c|c|c|c|c|c|c|c|c|c|}
\hline
Nuclear Mass &\multicolumn{3}{c|}{${m_1/m_{-1}}^{1/2}$ (MeV)} 
&\multicolumn{2}{c|}{$\Gamma$}\\ \hline
 &pairing+MEM&Our work&Expt.&our work & Expt.\\ \hline
$^{204}Pb$&13.4&13.6&13.7$\pm$0.1&2.02&3.3$\pm$0.2 \\
$^{206}Pb$&13.4&13.51&13.6$\pm$0.1&2.03& 2.8$\pm$0.2 \\
$^{208}Pb$&13.4&13.44&13.5$\pm$0.1&2.03&3.3$\pm$0.2  \\
\hline
\end{tabular}
\end{table}
\bigskip

In Table II, we have displayed the data obtained from a recent experiment
\cite {dpatel13} and compared our results. Column two of the table is also devoted 
to the result obtained from pairing plus MEM effect \cite{khan09}. The data 
show clearly that our result (extended Thomas-Fermi) has a priority over 
the pairing + MEM prediction. For example, the difference between the 
pairing+MEM results and experimental observation is 0.3 MeV for $^{204}$Pb 
isotopes, which is away from the experimental error, while it is 
only 0.1 MeV (within the error bar) in the RETF and data. 
This trend also followed by $^{206}$Pb and $^{208}$Pb nuclei. In our model,
we have not included any pairing externally. But still our results are 
good enough in comparision with MEM+pairing. This implies two things:
(i) pairing may not be impotant in calculation of excitation energy 
or (ii) pairing effect is automatically included in Thomas-Fermi 
calculations. 
\begin{figure}
\includegraphics[scale=0.31]{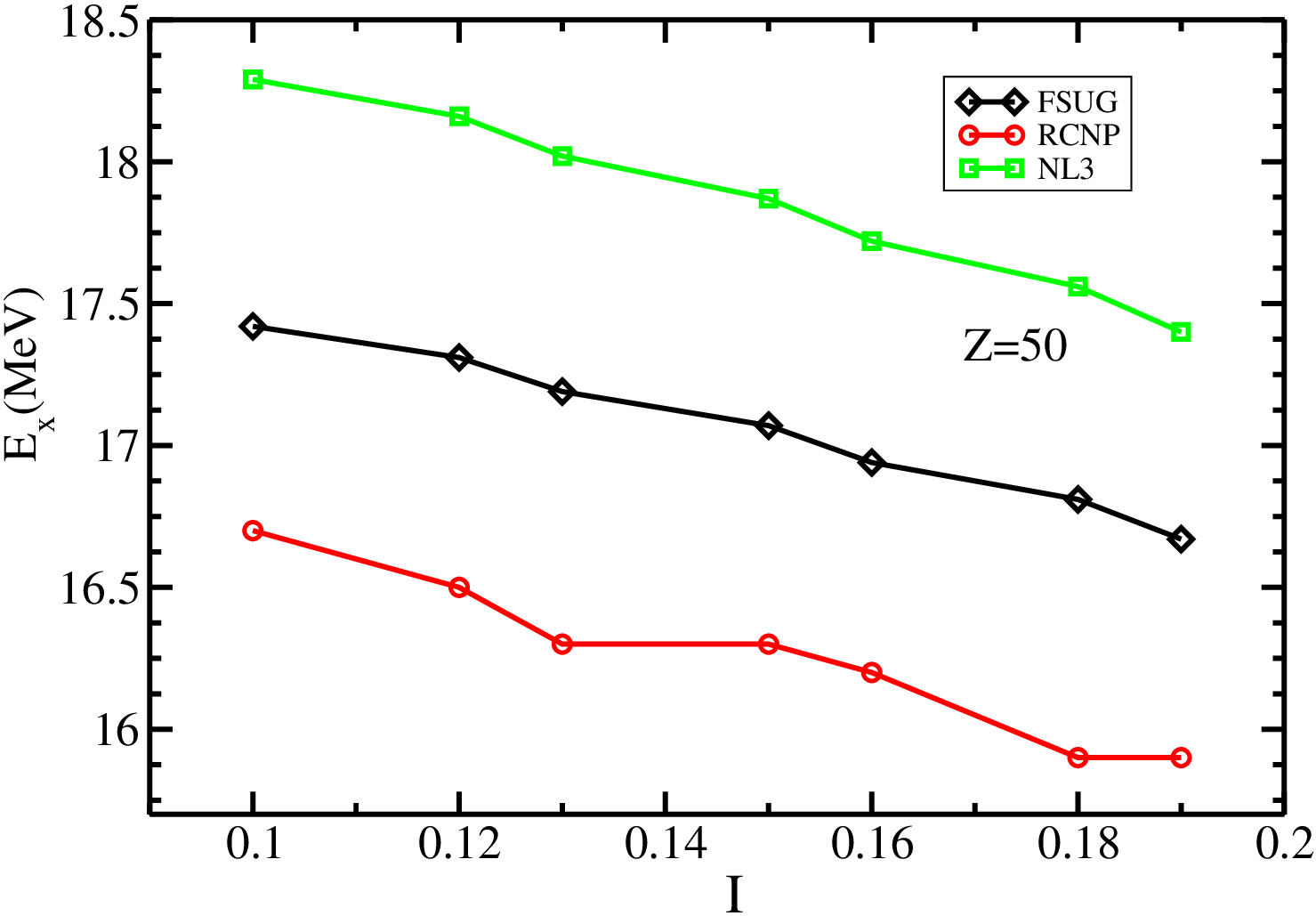}
\caption{\label{fig3} The variation of giant monopole excitation energy 
$E_X$ with proton-neutron asymmetry I=(N-Z)/(N+Z) for Sn isotopes.}
\end{figure}
To our understanding, the second option seems to be more 
appropiate, because lots of work show that pairing must be included for 
the calculation of excitation energy of open shell nuclei. Fig. 3 shows the
variation of excitation energy with proton-neutron asymetry in Sn 
isotopes. Here, we want to know, how the excitation energy varies with 
asymetry or more specifically, "is the variation of experimental 
excitation energy with asymetry same that of theoritical one ?". 
The graph shows that the variation with both NL1 and FSUG are 
following similar partten as experimental one with a
different mangnitued as shown in the figure.

\begin{figure}
\vspace{0.6cm}
\hspace{-0.3cm}
\includegraphics[scale=0.30]{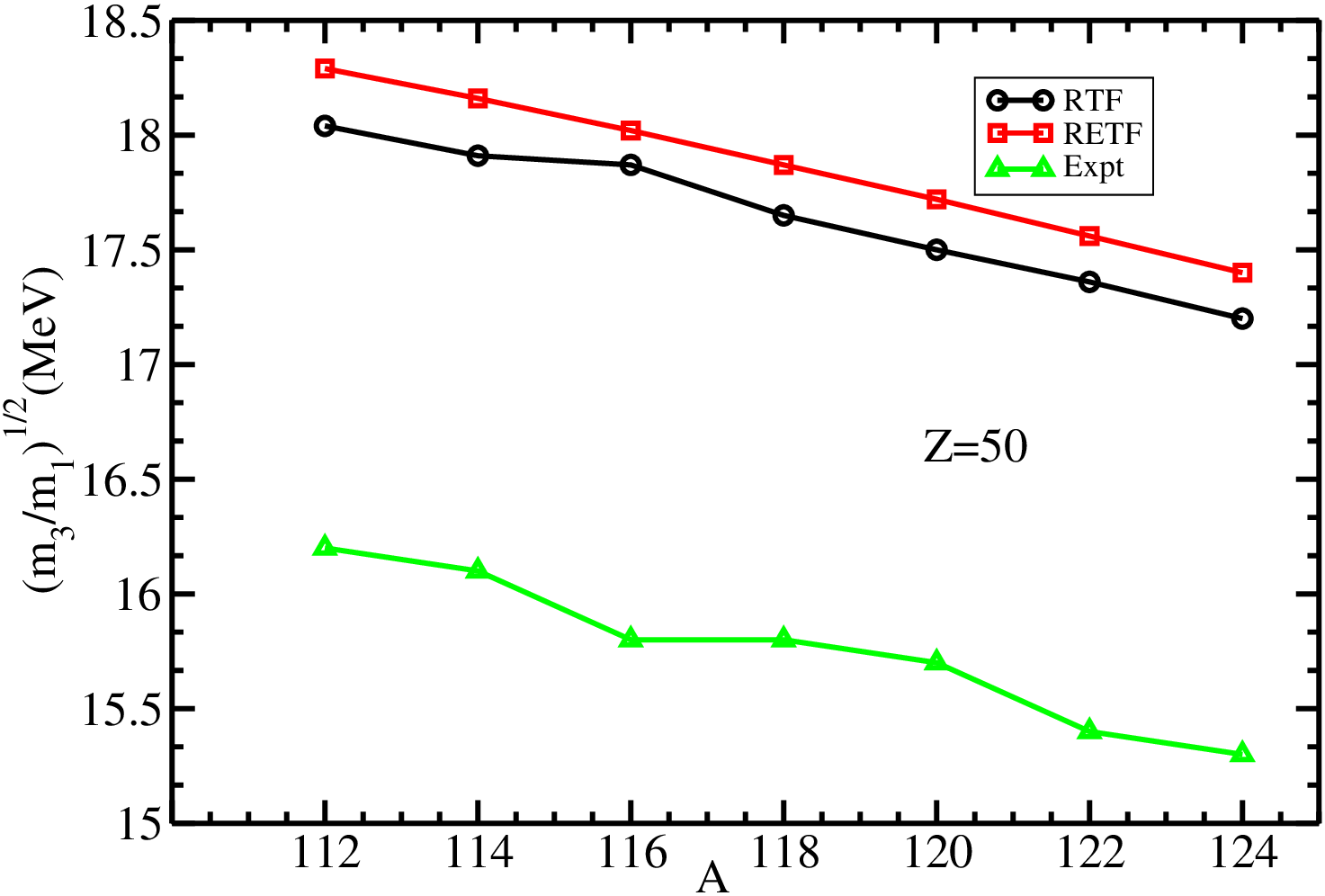}
\caption{\label{fig4} The scaling monopole excitation energy within  
RETF and RTF formalisms compared with the experimental momentum 
ratio ${m_3/m_1}^{1/2}$ \cite{dpatel13}.} 
\end{figure}

\begin{figure}
\vspace{0.6cm}
\hspace{-0.3cm}
\includegraphics[scale=0.32]{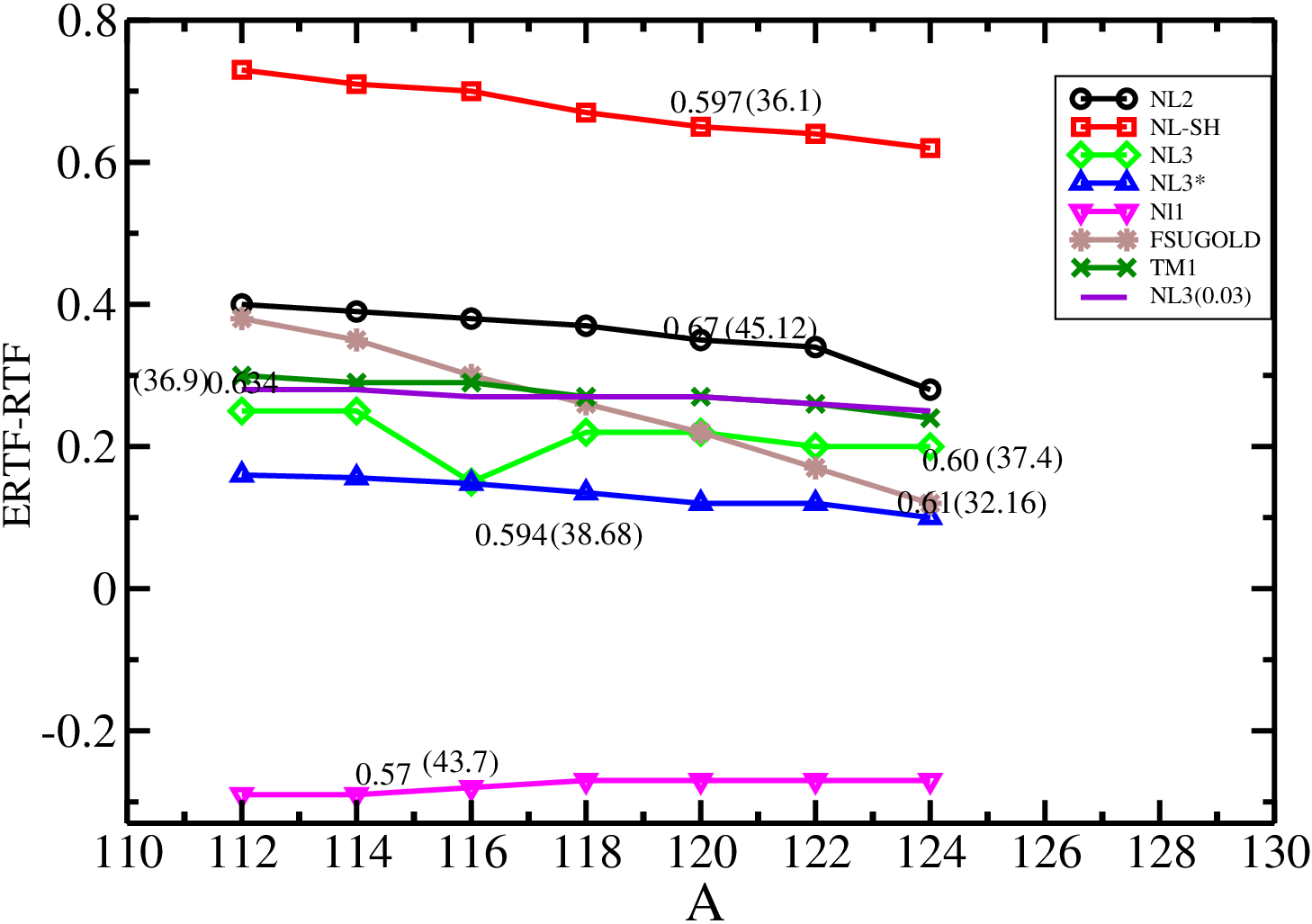}
\caption{\label{fig5} The variation of the difference 
of giant monopole excitation energy obtained from RETF and RTF 
(RETF-RTF) formalisms with various parameter sets for Sn isotopic chain. }
\end{figure}
In Figure 4,  we have compared the results obtained from
RETF and RTF with the experimental data for Sn isotopes. The celebraty NL3 
parameter set is used in the calculations.
The graph shows that there is only a small difference ($\sim 0.2$ MeV) in 
RETF and RTF results. Interestingly, the RETF correction is additive 
to the RTF result instead of softening the excitation energy 
of Sn isotopes. Then the natural question arises: is it the
brhavior for all the parameter sets in  RETF approximation ?. 
To attend the question, we plotted Fig. 5, where we have shown the
difference of ${\sqrt{m_3/m_1}}$ obtained with RETF and RTF results 
(RETF-RTF) for various parameter 
sets. For all sets, except NL1, we find  RETF-RTF as positive. Thus,
it is a challenging task to antangle the term which is the responsible
factor to determine the sign of RETF-RTF. Surprisingly, for most of 
the parameter sets, RTF is more towards experimental data. Inspite 
of this, one cannot says anything about the qualitative behavior of 
RETF. Because, the variation of the density at the 
surface taken care properly by RETF formalism, which is essential. 
One more interesting observation is that, when one investigate the 
variation of RETF-RTF in the isotopic chain of Sn, it remains almost constant 
for all the parameter sets, except FSUG. In this context, FSUG behaves 
differently.

Variation of RETF-RTF with neutron-proton asymmetry for FSUG set shows that,
there may be some correlation of RETF with the symmetry energy. 
This is clearly absent in all other parameter sets. Now it is essential 
to know, in which respect the FSUG parameter set is different from other. 
The one-to-one interaction terms for NL3, NL2, NL1 and NL-SH all 
have similar couplings. However, the FSUG is different from the 
above parameters in two aspect, i.e., two new coupling 
constants are added. One corresponds to the self-interaction of $\omega$ and
other one corresponds to the isoscalar-isovector meson coupling. 
It is known that self-interaction of $\omega$ is responsible for 
softening the EOS \cite {gmuca,toki94,bodmer91} and the isoscalar-isovector
coupling takes care of the softening for symmetry energy of 
symmetric nuclear matter\cite {horow02}. The unique behavior shown by the FSUG 
parametrization may be due to the following three reasons:
\begin{enumerate}
\item  introduction of isoscalar-isovector meson coupling $\Lambda_V$.
\item  introduction of self-coupling of $\omega-$meson. 
\item Or simultaneous introduction of both these two terms with refitting 
       of parameter set with new constraint.
\end{enumerate}
\begin{figure}
\vspace{0.6cm}
\hspace{-0.3cm}
\includegraphics[scale=0.32]{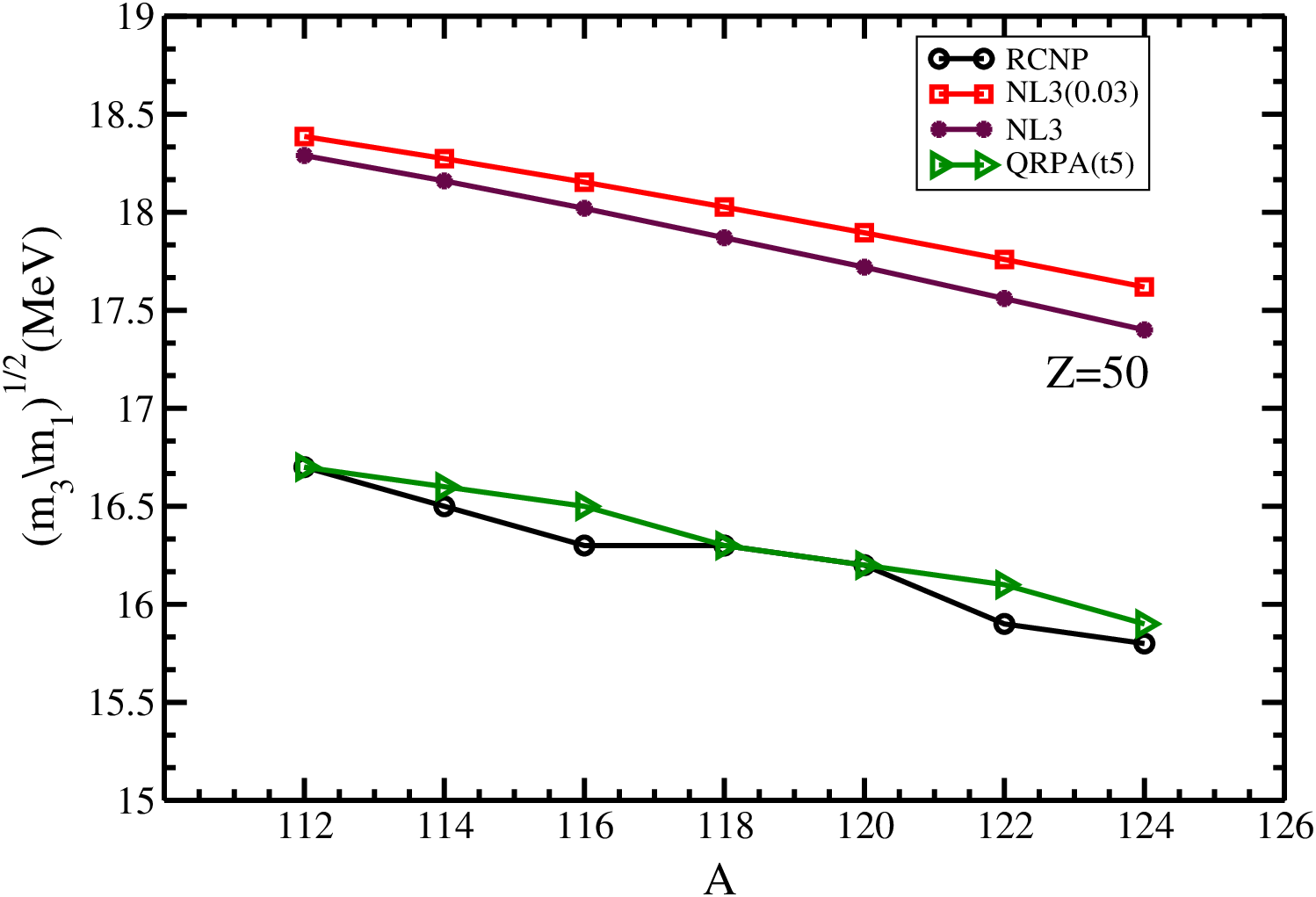}
\caption{\label{fig6} The momentum ratio ${\sqrt{{m_3/m_1}}}$ for Sn isotopes
obtained with NL3+$\Lambda_V$ is compared with NL3, QRPA(T5) and 
experimental data. }
\end{figure}
In order to discuss the first possibility, we plotted NL3+$\Lambda_V$(0.03) 
in Figure 6. The graph shows that there is no difference between NL3 
and NL3+$\Lambda$, except the later set predicts a more possitive 
RETF-RTF. It is well known that, the addition of $\Lambda_V$ coupling, i.e.,
NL3+$\Lambda_V$(0.03) gives of a softer symmetry energy 
\cite{patra9}. This implies that, models with softer energy have greater 
difference in RETF and RTF. At a particular proton-neutron asymmetry, 
RETF-RTF has a larger value for a model with softer symmetry energy. This 
observation is not conclusive, because all the parameter sets do not 
follow this type of behavior. Quantitatively, the change of RETF-RTF 
in the Sn isotopic series is about $70 \%$, while this is 
only $20-30\%$ in NL3 and other parameter set. 

In Table III, we have 
listed the $\rho-$meson contribution to the total energy. From the analysis of our 
results, we find that only $\rho-$contribution  to the total binding energy 
change much more than other quantity, when one goes from RTF to RETF. But 
this change is more prominent in FSUG parameter set than other sets 
like NL1, NL2, NL3 and NL-SH. Simple assumption says that, may be the absent 
of $\Lambda_V$ term in other parameter is the reason behind this. But we have 
checked for the parameter NL3+$\Lambda_V$, which does not follow. This 
also shows similar behavior like other sets. In Table IV, we have given 
the results for FSUG, NL3+$\Lambda_V$ and NL1. The data show clearly 
that, there  is a huge difference of monopole excitation energy 
in RETF and RTF with FSUG parameter set. 
For example, the $\rho-$meson contribution to the GMR in RETF for 
$^{112}$Sn is 21.85 MeV, while in RTF it is only 0.00467 MeV.  
\begin{table}
\caption{\label{tab:table1}{Contribution of the $\rho-$meson to the total
binding energy in the RTF and RETF approximations with FSUGold and NL1
parameter set.}}
\begin{tabular}{|c|c|c|c|c|c|c|c|c|c|c|}
\hline
Mass &\multicolumn{4}{c|}{FSUG} &\multicolumn{2}{c|}{NL3(0.03)}&
\multicolumn{2}{c|}{NL1}\\ \hline
&RETF&RTF&RETF$\Lambda_V$&RTF$\Lambda_V$ & RETF &RTF&RETF&RTF\\ \hline
112& 21.85&-6.66&-0.00130&0.00467&20.60 &20.12&17.91&16.73 \\
114& 28.72&-8.64&-0.00202&0.00664& 27.11&26.48&23.51&22.00  \\
116&36.42&-10.83&-0.00248&0.00829&34.40&33.62&29.73&27.90 \\
118&44.87&-13.23&-0.00298&0.01013&42.42&41.499&36.52&34.37 \\
120&54.04&-15.82&-0.00353&0.01213&51.13&50.05&43.84&41.37\\
122&63.88&-18.58&-0.00411&0.01429&60.48&59.24&51.63&48.85\\
124&74.33&-21.49&-0.00473&0.01660&70.43&69.03&59.84&56.76\\
\hline
\end{tabular}
\end{table}
However, this difference is nominal in NL3+$\Lambda_V$  parameter set,
i.e., it is only 0.48 MeV. Similarly, this value is 1.18 MeV in 
NL1 set.  The contribution of $\rho-$meson to total 
energy comes from two terms: (i)  one from $\Lambda_V{R^2}{V^2}$ and 
other (ii) from $\rho^2$. We have explicitly shown that contribution comes from 
$\Lambda_V{R^2}{V^2}$ makes a huge difference between the GMR 
obtained from RETF and RTF formalisms. This type of contribution does not 
appear from NL3+$\Lambda_V$.  For example, in $^{112}$Sn the contribution 
of $\Lambda_V{R^2}{V^2}$ with RETF formalism is -6.0878 MeV, 
while with RTF formalism is -5.055 MeV. 

The above discussion gives us a significant signiture that the contribution of $\Lambda_V$ may 
be responsible for this anomalous behavior. But an immediate question arises 
, why NL3+$\Lambda_V$ parameter set does not show such type of effects, 
inspite of having $\Lambda_V{R^2}{V^2}$ term. This may be due to 
the procedure in which $\Lambda_V{R^2}{V^2}$ term is added in two 
parameters. In NL3+$\Lambda_V$(0.03), the $\Lambda_V{R^2}{V^2}$ term 
is not added independently. The $\Lambda_V$ and $g_\rho$ are interdependent to 
each other to fix the binding energy BE and difference in neutron and proton 
rms radii $R_n$-$R_p$. But in FSUGold, $\Lambda_V$ coupling 
constant is added independenly to reproduce the nuclear observables. 
\begin{table}
\caption{\label{tab:table1}{$^{S}K_A$ and $^{C}K_A$ are compressibility of finite 
nuclei obtained from scaling and costraint methods, respectively are compared
with the values obtained from the equation of state (EOS). 
}}
\bigskip
\begin{tabular}{|c|c|c|c|c|c|c|c|c|}
\hline
Nuclear Mass &\multicolumn{3}{c|}{NL3} &\multicolumn{3}{c|}{FSUGOLD}\\ \hline
 &$^{S}K $&$^{C}k$&$^{EOS}K$ &$^{S}K$ & $^{C}K$& $^{EOS}K$ \\ \hline
 $^{208}Pb$& 164.11& 149.96&145 &147.37& 134.57&138.42\\
 $^{116}Sn$&164.64& 155.39&131.57 &147.11& 139.71&127.64 \\
$^{40}$P&136.70&110.43&105&123.40&100.36&102.53 \\
$^{40}$Ca&145.32&134.47&105 &130.93& 123.15&102.53 \\
\hline
\end{tabular}
\end{table}
In Table IV, we have listed the compressibility of some of the selected nuclei 
in scaling $^SK_A$ and 
constraint $^CK_A$ icalculations. This results are compared  with the computed values 
obtained from EOS model. 
To evaluate the compressibility from EOS, we have followed the procedure discussed 
in \cite{centelles09,shailesh13}.  M. Centelles et al \cite{centelles09},
parameterised the density for finite nucleus as $\rho_A=\rho_0-\rho_0/(1+c*{A^{1/3}})$ 
and obtained the asymmetry coefficient $a_{sym}$ of the nucleus with mass A 
from the EOS at this particular 
density. Here also, we have used the same parametric from of the density and obtained 
the compressibility of finite nucleus from the EOS. For example, $\rho_A=0.099$ for 
$A=208$ in FSUG parameter set. We have calculated the compressibility from 
the EOS at this particular density, which comes around 145 MeV. We have also 
calculated the compressibility independently in Thomas-Fermi and extended 
Thomas-Fermi using scaling and constraint calculations, which are 161 MeV and 146.1 MeV, 
respectively.

\section{Summary and Conclusion}
In summary, we analysed the predictive power of various force parameters, 
like NL1, NL2, NL3, Nl-SH and FSUG in the frame-work of relativistic 
Thomas-Fermi and relativistic extended Thomas-Fermi approaches for giant 
monopole excitation energy of Sn-isotopes. The calculation is then extended
to some other relevant nuclei. 
The analysis shows that Thomas-Fermi approximation gives better resluts than 
pairing+MEM data. It exactly reproduces the experimental data for Sn isotopes, when
the compressibility of the force parameter is within $210-230$ MeV.
We also concluded that a parameter set can reproduce the excitation energy 
of Sn isotopes, if its infinite nuclear matter compressibility lies within 
$210-230$ MeV, however, fails to reproduce the GMR data for other nuclei within
the same accuracy. 

We have qualitatively analized the difference in GMR energies RETF-RTF
using RETF and RTF formalisms in various force parameters. The FSUGold
parameter set shows different behavior from all other forces. Also, 
extended our calculations of monopole excitation energy for Sn isotopes
with a force parametrization having softer symmetry energy (NL3+ $\Lambda_V$). 
The excitation energy decreses with the increse of proton-neutron 
asymetry agreeing with the experimental trend. In conclusion, after 
all these thorough analysis, it seems that the softening of Sn isotopes 
is an open problem for nuclear theory and more work in this direction are
needed.

\end{document}